\documentclass[a4paper,fleqn]{cas-sc}

 \usepackage[numbers]{natbib}

\def\tsc#1{\csdef{#1}{\textsc{\lowercase{#1}}\xspace}}
\tsc{WGM}
\tsc{QE}
\tsc{EP}
\tsc{PMS}
\tsc{BEC}
\tsc{DE}

\begin{document}
\let\WriteBookmarks\relax
\def\floatpagepagefraction{1}
\def\textpagefraction{.001}

\shorttitle{Role of Humour in SE}


\title [mode = title]{The Role of Humour in Software Engineering - A Literature Review and Preliminary Taxonomy}                      




\author[1]{Dulaji Hidellaarachchi}

\cormark[1]

\fnmark[1]

\ead{dulaji.hidellaarachchi@rmit.edu.au}



\affiliation[1]{organization={ School of Computing Technologies, RMIT University},
   city={Melbourne},
     state={Victoria},
   country={Australia}}

\author[2]{John Grundy}
\affiliation[2]{organization={Faculty of Information Technology, Monash University},
   city={Melbourne},
     state={Victoria},
   country={Australia}}

\fnmark[2]
\author[2]{Rashina Hoda}
\affiliation[2]{organization={Faculty of Information Technology, Monash University},
   city={Melbourne},
     state={Victoria},
   country={Australia}}

\fnmark[2]







\begin{abstract}
Humour has long been recognized as a key factor in enhancing creativity, group effectiveness, and employee well-being across various domains. However, its occurrence and impact within software engineering (SE) teams remains under-explored. This paper introduces a comprehensive, literature review-based taxonomy exploring the characterisation and use of humour in SE teams, with the goal of boosting productivity, improving communication, and fostering a positive work environment while emphasising the responsible use of humour to mitigate its potential negative impacts. Drawing from a wide array of studies in psychology, sociology, and organizational behavior, our proposed framework categorizes humour into distinct theories, styles, models, and scales, offering SE professionals and researchers a structured approach to understanding humour in their work. This study also addresses the unique challenges of applying humour in SE, highlighting its potential benefits while acknowledging the need for further empirical validation in this context. Ultimately, our study aims to pave the way for more cohesive, creative, and psychologically supportive SE environments through the strategic use of humour.
\end{abstract}



\begin{keywords}
Humour\sep Software Engineering \sep Taxonomy \sep Human Aspects 
\end{keywords}

\maketitle

\section{Introduction}

The software engineering (SE) industry is characterized by high demands for precision, creativity, and intense collaboration among team members \cite{knobelsdorf2008creativity, boehm1988understanding}. In such a demanding environment, human aspects play a crucial role in the success of software development processes and their outcomes \cite{sommerville1996human, colomo2010study, rajeswari2003development, feldt2010links}. Researchers have increasingly highlighted the importance of integrating these human aspects into SE, arguing that this integration is essential for building complex software systems with desired qualities \cite{murphy2010human}. Various studies have explored diverse human aspects, such as personality \cite{cruz2015forty, capretz2010we}, emotions \cite{sanchez2019taking, miller2015emotion}, gender \cite{carver2019gender, burnett2016gendermag}, and empathy \cite{akgun2015antecedents, gunatilake2023empathy}, and etc. across different SE contexts \cite{wang2019implicit, hannay2009effects, hidellaarachchi2024impact, kanij2013empirical}. A key component of these human aspects is team dynamics and employee well-being, which are critical to the success and productivity of software projects \cite{restrepo2024characterizing, hoffmann2022human}. Effective team dynamics can enhance collaboration, innovation, and overall job satisfaction. Therefore, it is crucial to pay attention to these human aspects alongside technical skills within software teams, as people and technology go hand in hand. 

Humour is one such human aspect, which to date has been under explored in the SE team context. Humour is a multi-disciplinary field of research where researchers have conducted diverse studies in psychology \cite{abel1998interaction, cann2009positive, bitterly2020sarcasm, martin2003individual}, organizational management \cite{romero2008humor, robert2012wheel, pham2021laughing, watson2017humour, plester2009healthy, al2012contextual}, education \cite{barros2017use, christman2018instructor}, finance \cite{bergeron2008effects}, creativity \& innovation \cite{biemans2024so, khan2022leading} along with very limited studies in SE, Computing and Information Technology (IT) \cite{tiwari2024great, amjed2016effect, moussawi2021effect, krienke2017information}. Among these, humour has been predominantly studied in the management, psychology and education contexts. 

In a team working environment, humour is widely recognised for its potential to enhance communication, leadership, foster creativity \& innovation, increase team effectiveness, productivity along with fostering a sense of community within people in organizations. However, humour is also considered to be a double-edged sword -- it can have both positive but also negative impact \cite{malone1980humor}. Several studies emphasize the importance of appropriate humour, as it is not always viewed as a positive aspect of teamwork. Humour that goes too far over the line of appropriate behaviour may lead to negative impacts such as tension, conflicts and reduced team effectiveness within an organizational setting \cite{robert2012wheel, abel1998interaction}. 

In social science and psychology research, humour is considered as a medium which leads to psychological health, healthy relationships, and quality life as it acts as a moderator on reducing stress, anxiety, depression and improving coping strategies considering the gender differences, communities and cultures  \cite{pandin2023influence}. In education research, humour is considered as a pedagogical tool that can be used specifically in higher education in variety of disciplines \cite{bakar2018use, powell1985humour}. Researchers have studied the influence of using humour in teaching mathematics \cite{amran2022use, menezes2021humour}, physics \cite{berge2017role}, medical teaching \cite{ziegler1998use}, non-fiction writing \cite{hogue2011m}, software engineering \cite{barros2017use}, economics \cite{wyk2011use} and even outdoor education \cite{hoad2013potential}. These studies highlight that the use of humour can be a trigger for increasing students' emotional engagement with the subject topics and classroom activities related to student-student, student-teacher and individual-context learning interactions. 

Despite the extensive research on humour in these fields, its role in the SE domain remains much less explored. SE is traditionally perceived as a technical and discipline, where human and social dynamics often receive less attention compared to more technical challenges \cite{tiwari2024great, sharp2009role}. However, as the complexity of software projects increases and the emphasis on team collaboration grows, understanding how humour can be leveraged positively to enhance team dynamics and project outcomes would be an added advantage.
\cite{graziotin2015affect}. Among the limited studies in SE/ IT, humour impact has been explored in user perceptions of personal intelligent agents like Siri and Alexa \cite{moussawi2021effect}, its influence on creativity and transformational leadership among employees in software organizations \cite{amjed2016effect, khan2022leading}, enhancing developer engagement through humour in SE \cite{tiwari2024great}, its role in requirements elicitation in information systems design \cite{krienke2017information}, and its application in software engineering education \cite{barros2017use}. These studies highlight the need for more research focusing on humour in diverse SE areas considering its potential impact. For instance, incorporating humour into software teams and practices can not only lead to notable improvements in team dynamics and leadership effectiveness, but also in SE activities such as requirements elicitation, designing, code reviews \cite{tiwari2024great}. Humour has the potential to reduce stress during intensive coding sessions or to make tedious tasks like documentation more engaging, which could significantly influence the overall performance and the quality of software projects \cite{tiwari2024great}. It also has the potential to reduce the cognitive load during complex problem-solving tasks, enhancing the creative and innovative software solutions \cite{krienke2017information}.

Despite these potential impacts, there is a distinct lack of comprehensive research that examines how humour affects people involved in SE, practices and overall projects. In this paper, we seek to address this gap by reviewing and synthesizing the existing literature on humour within organizational and software engineering contexts.  \textcolor{black}{Our aim is to bridge this gap of in SE research by conducting a narrative literature review within SE and other disciplines by drawing insights from existing literature on humour, identify opportunities for characterising potential positive and negative types and uses of humour potentially relevant to SE teams and contexts and providing a preliminary taxonomy of humour for the exploration of future studies of the impact of humour on SE. This taxonomy integrates humour theories, styles, models and scales from various disciplines, providing a structured framework, which will eventually serves as a foundation for future research to explore the role of humour in SE and guiding practitioners in its application. In the sections that follow, we first review the existing literature on the role of humour in SE and beyond (section \ref{related work}). Then we present our preliminary taxonomy of humour along with how we developed it (section \ref{methodology} and \ref{taxonomy}), and a discussion on key theories and scales that can be adopted to assess humour in the SE context, limitations and directions for future research (section \ref{discussion}). This will conclude with a summary of our research (section \ref{conclusion}).}

\section{Related Work} \label{related work}
\subsection{Dynamics of Humour} \label{dynamics of humour}

Humour is considered as a basic element of human interaction and is considered to be a complicated, multifaceted phenomenon that is highly subjective. \textcolor{black}{It does not have a single standard definition \cite{robert2007case}. Among the diverse set of definitions, we consider humour as \emph{``an ubiquitous human interpersonal behaviour, consists of amusing communications that produce positive emotions and cognition in the individual, group or organisation "}, one of the most referred definitions in the literature \cite{romero2006use}. In simple terms, it can also be described as \emph{``humorous events that promote laughter and pleasure"} \cite{plester2009crossing}. }Further, researchers have also considered humour as a stable personality trait or individual difference variable that is conceptualised as a cognitive ability \cite{feingold1993preliminary}, an aesthetic response \cite{ruch1998two}, an habitual behaviour patter \cite{craik1996sense, ruch2007temperament}, an emotion-related temperament trait \cite{ruch1996assessing}, an attitude \cite{svebak1996development}, a coping strategy/ tool or defense mechanism \cite{svebak1996development} and so on. 

From another perspective, psychologists have defined humour as \emph{``a normal verbal conduit of communication whereby there is a sender and a receiver with the underpinning technicalities of encoding, noise, and decoding} \cite{al2012contextual}. As a result of this diverse nature of humour and the broad scope of humour research, it is challenging to highlight all current research in the area. Additionally, terms such as laughter, sarcasm, and jokes are often associated with humour and used in similar contexts \cite{bitterly2020sarcasm}. However, studies also emphasised that although these terms are closely related, they are certainly not the same \cite{mulder2002humour}. For instance, laughter may be an outcome of humour, but not all humorous situations result in laughter. Likewise, jokes are commonly considered as the prototypical form of verbal humour \cite{dynel2009beyond}. Hence, understanding these nuances is essential for advancing research in the dynamics of humour.

Prior studies also emphasize that not all humour has a positive impact and it can lead to negative impacts, such as destroying relationships depending on the context and the type of humour present \cite{cann2009positive}. When handled with care and in a responsible manner, humour has useful effects on individuals, groups and organizations and is worthy for further investigations \cite{plester2009healthy}. Among these, researchers have conducted studies on humour across diverse contexts, each resulted in insights into its role and impact. In organizational settings, humour has been investigated as tool to enhance workplace culture \cite{plester2009crossing, plester2009healthy}, improving group effectiveness \cite{romero2008humor}, employee well-being \cite{robert2012wheel}, decision-making \cite{watson2017humour},  communication \cite{cann2009positive} and emerging leadership \cite{amjed2016effect}. These studies have shown that humour can be helpful in reducing work stress, foster creativity \& innovation and enable employees to cope with their teams and come out of challenging working environments successfully \cite{pham2021laughing}.  Humour has been considered a lot in research in education where it has been examined for its potential to enhance student engagements and learning outcomes, Research suggests that humour can support more engaging learning environment, help in the retention of information and create a more positive classroom atmosphere while encouraging more student participation \cite{barros2017use, christman2018instructor, pandin2023influence}. 

Healthcare is another area where humour has been investigated. Here it has been studied as a therapeutic tool, particularly in patient care. It has been found out that healthcare professionals use humour as a coping mechanism to manage the emotional demands of their work and improve their relationships with patients \cite{de2020humor, haydon2015narrative, mccreaddie2014humour}. In  social science research, humour has been mainly studied as a individual trait that is helpful in communication, people's relationships building, diffuse conflicts and navigate social hierarchies \cite{abel1998interaction, goldstein2013cross}. Additionally, humour has been studied in media and entertainment where it is considered as a central element in diverse content such as television, social media, and movies \cite{chan2011selling, de2022case}. Research in this area often focuses on the effects of humour on audiences, such as its role in shaping public opinion, reinforcing stereotypes, or challenging societal norms. Compared to these, research on humour impact on SE us very limited and mostly focused on management aspect in the IT industry. Across these diverse contexts, humour is recognized not only for its positive effects but also for its potential drawbacks.  As a result, the study of humour continues to be a rich and dynamic field, contributing to our understanding of human behavior and interaction. In section \ref{humour SE} and \ref{humour other} we highlight some of the prominent research in these areas. 

\subsection{Humour Research in the SE Context} \label{humour SE}

Despite of increasing interest towards human aspects and their impact on SE, there has been very limited research on the influence of humour on SE teams and outcomes. Nevertheless, we identified a few studies that have explored humour in related areas such as  computing, SE education, information systems, human-computer interaction areas, requirements engineering. However, most of these studies focused on organizational, management and leadership aspects in the IT industry.

In \cite{amjed2016effect}, researchers examined the impact of humour on employee creativity and the role of transformational leadership among employees at two software development companies in Pakistan. They conducted a survey with 120 employees, using questionnaires to assess humour and leadership styles. The study found that positive types of humour enhance employee creativity, while negative humour styles have mixed effects. However, the research primarily focused on managerial aspects and emphasized that humour has a positive influence when supported by effective leadership. They also found that excessive use of humour can lead to negative outcomes. Although the study involved full-time employees from two software houses in Pakistan, it did not clearly specify the participants' roles, such as whether they were software developers, managers, or held other positions.

A multi-disciplinary study \cite{plester2009crossing} explored the boundaries of workplace humour and fun through an ethnographic study conducted in four New Zealand organizations across different industries. These included a small IT service company, two large utility organizations, and a medium-sized law firm. The study took a broad view of how workplace humour and fun are shaped by varying organizational cultures and boundaries. Through 59 semi-structured interviews, observations, and documentary data collection, researchers found that the formality level of each organization significantly influenced humour and fun.
In the IT service organization, which was predominantly male-dominated, the culture was more informal, with frequent banter among colleagues. The managing director of this company was often the initiator of humour. This was in contrast to the other three organizations, which had higher levels of formality and more restricted boundaries for humour. Additionally, the IT service company had a flatter career hierarchy compared to the others, which the researchers identified as a key reason for selecting this particular organization. They noted that the lively humour in the workplace contributed to enhancing employee expertise and fostering innovative IT solutions in the local industry. However, they also pointed out that male dominance in the organization might have contributed to the broader rules and boundaries for humour compared to the other organizations. 

In a further study \cite{moussawi2021effect}, researchers explored the impact of humour and the voice of personal intelligent agents (PIAs) like Siri and Alexa on users' perceptions, and how these perceptions relate to cognitive- and emotion-based trust. They conducted an online experiment with 271 undergraduate students in the USA, using a 2x2 factorial design with two levels of humour (humour and no humour) and two levels of voice (voice and text, text only). This study introduced a theoretical model that explains the effect of humour on user perceptions related to PIAs. The researchers found that, in the era of AI, for an AI system to generate original humour, it must effectively process information by producing diverse interpretations of linguistic expressions and selecting the most appropriate one for the context. Their findings highlighted that humour is an effective form of communication not only between humans but also between users and AI systems. However, they also emphasized the need to extend the study to include a more diverse user population that better represents typical PIA users. Additionally, they called for more research into how AI systems can recognize and respond to humour as part of their information processing capabilities.

Khan et al. \cite{khan2022leading} recruited 111 managers working in software houses in Pakistan to investigate the impact of leaders' humour styles on employee creativity. They used a Likert scale questionnaire to measure both humour styles and creativity among employees. The study found that the affiliative humour style (see Section \ref{humour assessment}) used by managers had a positive and significant effect on employee creativity. The authors suggested that future research should explore humour, along with other traits like individual differences, specifically within software development teams.
Mulder and Nijholt \cite{mulder2002humour} discussed the state of the art in humour research, particularly within the context of computer science, compared to other fields. They argued that developing computational models of humour would be beneficial for interface designers looking to incorporate humour into human-computer interaction. By emphasizing the multidisciplinary nature of humour, the study highlighted the need for more theoretical and empirical research on humour in the field of computer science, which could enhance our understanding of its concepts and applications within this context. \textcolor{black}{In their study Borsotti and Bjørn \cite{borsotti2022humor} explored humour within computer science organizations, focusing on how humour perpetuates stereotypes and creates barriers to inclusion. Through an equity-focused lens, they examined humour embedded in organizational artifacts, rituals, and traditions, highlighting how negative stereotypes, activated through humour, can trigger stereotype threat and reinforce exclusionary norms. The study emphasizes that humour, while often seen as harmless, can normalize biases and influence perceptions of who belongs in computing disciplines.  This research broadens the understanding of humour’s role in SE by linking it to diversity, inclusion, and organizational culture—topics often under-explored in traditional SE studies. However, the study primarily focuses on institutional practices rather than team-level interactions or humour's direct impact on SE. Further in \cite{friedman2002computer, friedman2003framework}, the authors discussed the concept of Computer-Oriented Humour (COHUM), describing it as `I-get-it' humour that relies on specialized computing knowledge. Through qualitative analysis of jokes, cartoons, and anecdotes, they identified COHUM as a tool for fostering in-group identity within computing communities. The study highlights how COHUM reflects the unique culture of computing professionals but notes its exclusivity to those with technical knowledge. The authors call for empirical research to explore how COHUM influences team cohesion and productivity within SE contexts.
}

Another study, examining three prominent examples of real-world humour in open-source software, explored how developers engage with SE through humour using  a case study \cite{tiwari2024great}. The researchers surveyed 125 software developers to investigate their personal practices and experiences with humour in the software they create. The case studies revealed that humour is present at every stage of the development process in open-source software, including coding, testing, committing, and integrating. They found that humour helps foster collaboration in open-source projects, extending beyond company boundaries. The survey found that over 93\% of respondents reacted positively to the presence of humour in software development. However, the study also noted that some respondents preferred to avoid incorporating humour in their software code, believing it could negatively impact its quality. The study primarily focused on humour in software code, specifically during testing and code commits, and did not address the dynamic nature of humour or its potential drawbacks. It lacks insights into crucial aspects such as responsible use of humour in SE, diverse activities and methodologies  and its effects on productivity, code quality, and software team performance. \textcolor{black}{Kuutila et al. \cite{kuutila2024makes} conducted an extensive analysis of programming-related humour by examining ~13K submissions from the r/ProgrammerHumor sub-reddit. Their study aimed to identify factors contributing to humour appreciation among software developers. Despite the complexity of predicting humour—evidenced by their regression models explaining only 10\% of the variance, their findings suggest that humour aligning with superiority and incongruity theories resonates more within the programming community. This study offers valuable insights into the cultural nuances of humour among software developers, highlighting the challenges of using NLP to predict humour appreciation}

Krienke and Bansal \cite{krienke2017information} conducted a literature review within the context of information systems (IS) design, specifically focusing on the role of humour in requirements elicitation. They provided a comprehensive overview of the existing literature related to the use of humour and its relationship to business strategy. The study highlights that, although current research suggests that the use of positive humour can be beneficial in obtaining high-quality and accurate requirements, there is little to no guidance on how to effectively implement humour in this context or provide recommendations to IT managers.
Based on the limited research available, the authors offered a set of instructions for IT managers on using humour during the requirements elicitation process and outlined possible future research directions in IT, project management, and IS concerning humour incorporation. They also emphasized the need for foundational work on the potential applications of existing humour-related theories, models, and frameworks to provide a comprehensive understanding for researchers interested in exploring this area further.

Barros et al. \cite{barros2017use} conducted a study on the use of comic strips in teaching software engineering, specifically focusing on software requirements specification. They conducted two controlled experiments and developed a guide to determine whether IT professionals and students could effectively specify software requirements using comic strips. The study emphasized that using comic strips made the requirements specification process more enjoyable. The researchers primarily conducted the study with computer science undergraduates and initially recruited a limited number of IT professionals to validate the understandability of the comic strips in the context of requirements. However, the exact number of IT professionals involved was not specified. The study also highlights the limited research in using humour as a tool in SE education, despite the broader body of research on humour in educational contexts. 

Although there is a growing recognition of the importance of human aspects in SE, the specific impact of humour on SE processes remains under-explored. While some research has touched on humour in related areas such as information systems, SE education, and human-computer interaction, the direct effects of humour on software development practices, team dynamics, and overall project outcomes have not been thoroughly investigated. Most of these existing studies focus on humour's role within organizational or managerial contexts, leaving a critical gap in understanding how humour can be systematically integrated into SE to enhance creativity, collaboration, and code quality. Filling this gap is essential, as it could lead to more innovative and effective approaches in software development, ultimately advancing the discipline.

\subsection{Humour Research in Other Contexts} \label{humour other}
There is considerable research on humour in non-SE contexts such as management, psychology, sociology, healthcare, and education. Humour has been extensively studied in relation to organizational culture, often referred to as workplace humour. These studies have explored the impact of humour on various aspects of organizational culture. For instance, in \cite{romero2008humor}, researchers analysed the literature to identify the relationship between organizational humour and group effectiveness. They developed the "Group Humour Effectiveness Model," which demonstrates how humour can improve communication, reduce stress, and foster a collaborative atmosphere within teams. Their model suggests that humour acts as a social lubricant, facilitating smoother interactions and stronger bonds among team members—factors crucial for effective teamwork in any organization. 

In another study \cite{robert2012wheel}, the impact of humour on employee well-being and happiness was discussed through a review of literature on humour, emotions, and well-being within the management field. The researchers developed a conceptual model known as the "wheel model of humour," which proposes that humour-induced positive affect can transmit emotions to social groups, creating a climate that supports the use of humour and subsequent humour events. This model focuses on organic or conversational humour, which is the most frequent form of organizational humour, rather than verbal jokes or cartoons. However, both of these models are conceptual and were developed based on literature reviews. They primarily discuss the effectiveness of humour in communication and leadership, with a focus on managerial roles. The models do not address the diversity of industries or domains, and they consider organizational culture as a whole, without highlighting specific industries or domains.

Pham et al. \cite{pham2021laughing} explored the role of humour in enhancing the effectiveness of meetings within organizations. By conducting an online survey that included a humour climate questionnaire—a scale used to assess humour (see Section \ref{humour assessment})—they found that both positive and negative humour, as well as playfulness, significantly impact meeting satisfaction and effectiveness. The study specifically focused on the complexities of negative humour and discovered that gender has a significant negative relationship with the negative aspects of humour. The authors emphasized the need for more empirical studies across diverse domains, as factors such as the type of meeting (virtual vs. in-person) and organizational culture could influence these findings. Humour has long been a significant area of research in psychology and sociology, with numerous empirical studies examining how it influences interpersonal dynamics, including communication, social relationships, interactions, and psychological well-being. Martin et al. \cite{martin2003individual} developed and validated a questionnaire to assess humour, aiming to explore its relationship with psychological well-being, social relatedness, and other aspects of individual differences. Similarly, in \cite{aslam2021effect}, researchers collected data from media houses in Pakistan via online questionnaire to investigate the role of humour in mediating workplace bullying. Both studies identified humour as a significant factor in enhancing psychological well-being and suggested further exploration across different countries, cultures, and domains.

Humour has also been explored as a tool in educational research, with numerous studies examining its influence across various educational settings, including schools, higher education institutions (colleges/universities), and different fields such as medicine, science, technology, engineering and math (STEM), economics, and physics. Menezes \cite{menezes2021humour} and Amran \cite{amran2022use} studied the impact of humour in the context of teaching mathematics in schools using mixed-method approaches. By involving both teachers and students, these studies provided diverse insights into the use of humour in teaching mathematics. From the teachers' perspective, humour was seen as highly relevant for teaching complex subjects like mathematics, as it helps reduce math anxiety. However, from the students' perspective, while they appreciated teachers using humour, they also noted gaps between their preferences and how humour was actually used in the classroom.

In higher education, studies by Powell \cite{powell1985humour}, Christman \cite{christman2018instructor}, Pandin \cite{pandin2023influence}, and Bakar \cite{bakar2018use} focused on the impact of humour on student engagement, resilience, and its potential as a pedagogical tool in college settings. These studies found that humour can effectively improve students' attention and interest, but it must be implemented responsibly in the classroom. They also emphasized the importance of providing proper training for teaching staff on how to use humour appropriately, ensuring the correct balance is maintained during classes. Hu et al. \cite{hu2017humour} explored the application of humour in STEM education, discussing its potential to enhance students' educational experiences. The study emphasized that humour can be beneficial, but only under certain conditions where it does not overwhelm students' cognitive load. By drawing on examples from the literature, the study provided insights into how humour can be effectively integrated into problem-solving approaches within STEM education.

Healthcare is another sector where humour has been extensively studied, particularly in relation to medical teaching, patient care, and interactions among healthcare professionals. Ziegler \cite{ziegler1998use} conducted a study on the use of humour in medical teaching, discussing the importance of incorporating humour at every stage of medical education. The author provided strong support from the literature, highlighting 18 claims that emphasize the significant positive impact of using humour in teaching, including in medical education. The study also outlined various roles of humour in medicine, such as patient and public education, relieving tension in clinical settings, dealing with uncertainty in clinical decisions, and dispelling myths.
In another study \cite{de2020humor}, the importance of humour was examined from the perspectives of doctors and nurses. This qualitative study involved 88 healthcare professionals and found that humour plays a crucial role in therapeutic relationships. However, it must be used in moderation, with careful consideration of patients' socio-cultural backgrounds before employing it.
Additionally, in \cite{mccreaddie2014humour}, the use of humour was explored from the patients' perspective through a constructivist grounded theory study. The research revealed that patients generally appreciate the use of humour and expressed a desire for healthcare professionals to initiate and reciprocate humour. However, the study also emphasized that initiating humour involves risks, requiring a certain level of self-esteem and confidence from healthcare professionals, but ultimately, it is a risk worth taking.

\subsection{Humour Assessment} \label{humour assessment}

Considering the multifaceted nature of humour and its application across diverse contexts, assessing humour is recognized as a complex process. Numerous theories, models, frameworks, styles, and scales have been developed to evaluate humour, each offering different approaches to capture its nuances. For instance, theories such as the \emph{superiority theory} emphasize the social dynamics and power relations embedded in humorous interactions while the theories, such as \emph{the relief theory}, explore humour as a means of releasing psychological tension \cite{martin2003individual, mulder2002humour}. These theories have informed the development of various models and frameworks that aim to systematically assess humour, including models that examine individual differences in humour appreciation, and analyze humour within specific cultural or organizational contexts \cite{ robert2012wheel, al2012contextual, ruch1998two}. In educational settings, frameworks have been developed to understand how humour can be used to enhance learning outcomes, reflecting the diverse applications of humour assessment \cite{ziv1988teaching,wanzer2010explanation}.

Researchers have also examined the assessment of humour through different styles and scales, reflecting the complexity and diversity of humour across cultures and contexts. The Humour Styles Questionnaire (HSQ), for example, is a widely used scale that categorizes humour into four styles: affiliative, self-enhancing, aggressive, and self-defeating \cite{martin2003individual}. The Multidimensional Sense of Humour Scale (MSHS) and the Coping Humour Scale (CHS) are other examples that assess how humour is used as a coping mechanism and its impact on mental health and well-being \cite{thorsonmultidimensional, martin1996situational}. These scales have been instrumental in understanding the role of humour in various contexts, including its potential benefits and drawbacks. Despite these measures, the assessment of humour remains a challenging and evolving field (due to its multidimensional nature), with ongoing debates about the best methods to capture its subtleties and the contextual factors that influence its effectiveness.

However, within the SE context, the application of these humour assessment scales remains largely unexplored. Notably, only two studies conducted in Pakistan have utilized the Humour Styles Questionnaire (HSQ) to examine the influence of humour in SE contexts, focusing on creativity and transformational leadership within software organizations \cite{amjed2016effect, khan2022leading} . This limited use highlights the gap in comprehensive humour assessment within SE, where the complex and technical nature of the field does not lend itself easily to traditional humour assessment models. This lack of extensive research leads us to develop a preliminary taxonomy of humour which is crucial for understanding how humour can be systematically assessed and leveraged to improve team dynamics, enhance creativity, and foster a positive work environment in SE settings. By developing this preliminary taxonomy, our goal is to provide a comprehensive understanding and possible application of humour in the SE context where researchers and practitioners can better harness the benefits of humour in this highly collaborative SE field.

\section{Development of Our Preliminary Taxonomy} \label{methodology}

While various theories, models, frameworks, and scales have been developed and applied in many fields, indicating the significance of humour, there is a notable lack of humour research within SE. We observe that there is no clear taxonomy to distinguish or make sense of the dynamic nature of humour in SE teams, leading to a limited understanding of its role and potential in SE.
To address this gap, we have developed a preliminary taxonomy aimed at providing a comprehensive understanding of how the dynamic nature of humour can be investigated in SE contexts. Our taxonomy also seeks to identify which existing theories, models, and scales might be applicable in SE research. By doing so, we aim to make this knowledge more accessible to SE researchers, encouraging further exploration of humour's impact within the SE community—an area that remains under-explored but may hold significant potential for advancing the field.

To develop our preliminary taxonomy of humour, we conducted an extensive analysis of existing research across multiple disciplines. We conducted a narrative literature review to develop our preliminary taxonomy. Unlike a systematic literature review (SLR), which follows a structured and rigorous methodology for selecting, analyzing, and synthesizing papers, we selected the narrative approach. Our focus was to conduct a preliminary review and analysis of key humour-related work, highlighting this under-researched aspect in SE research while exploring trends and new insights. We specifically examined studies discussing humour in professional settings, emphasizing its impact on teamwork, communication, and collaboration within SE and related fields. The reviewed literature spans multiple disciplines, including psychology, management, organizational behavior, and SE. We identified literature primarily through an exploratory review of Google Scholar, which served as our primary source for relevant papers. To ensure broad coverage, we selected studies from diverse domains that discussed SE or related IT areas, including: IT organization management, information systems, computing and IT management etc, along with psychology, management and social science context which we have identified as domains that humour has been majorly considered in professional settings. Given the scarcity of studies directly examining humour in SE, we also used backward and forward snowballing techniques to locate additional relevant studies. Our primary search terms included \emph{``humour", ``software engineering", and ``organization \& management"}. For each selected paper, we extracted data using predefined guiding questions, such as:
\begin{itemize}
    \item What type of humour is considered in the paper?
    \item How do the authors define humour?
\item What impact is identified?
\item What is the domain of the study?
\item What are the key objectives, findings, limitations, and future work?
\end{itemize}
These insights allowed us to categorize humour-related findings based on recurring themes and existing classifications, forming the foundation of our taxonomy. Instead of using formal thematic analysis or coding schemes, we refined the taxonomy through multiple iterations, ensuring coherence across disciplines while maintaining relevance to software engineering. \textcolor{black}{The first author led the initial extraction and categorization resulted in a working draft of the taxonomy which was formed by manually grouping recurring patterns and concepts into thematic clusters. After that all authors engaged in iterative discussions to group and structure the components under the four key dimensions: theories, styles, models/frameworks, and scales. During this process some categories were refined (e.g. separated ``models" from ``scales") while these discussions helped ensure conceptual clarity, eliminate redundancy, and maintain relevance to SE contexts. At this point, we had reviewed a total of 28 papers (see the list of papers in table \ref{TABLE: paper list}), which we considered as sufficient for the scope of our preliminary taxonomy. This taxonomy is intended as a conceptual foundation and by following these steps the study can be replicated, extended to improve its comprehensiveness, empirically validated and operationalised within the SE context.} 

\section{The Preliminary Taxonomy of Humour} \label{taxonomy}
\begin{figure}[htbp]
  \includegraphics[width=1.05\linewidth]{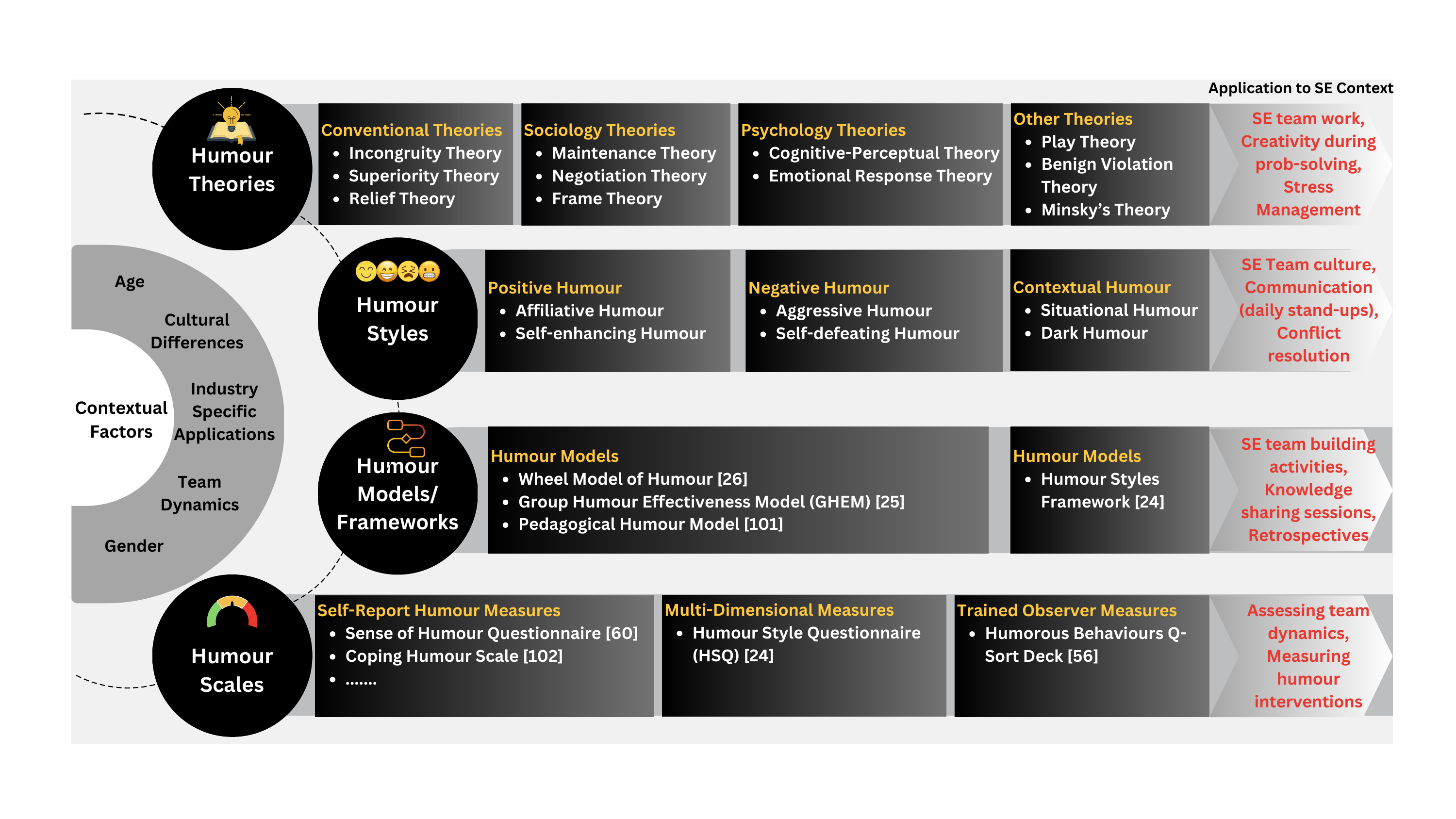}
  \caption{A Preliminary Taxonomy of Humour}
  \label{Taxonomy}
\end{figure}

Through the above-mentioned process, we identified that humour has been studied and categorized in various domains based on its nature, formation, and usage. \textcolor{black}{As a result, we identified four distinct groups of humour which we have used to develop and structure our proposed taxonomy.  These groups are: humour theories, humour styles, humour models/frameworks, and humour scales, each group represents a key aspect of humour research. Our aim was to create a structured and comprehensive taxonomy that organizes existing humour research while making it relevant to SE context.}
 While these groups are presented as disjoint for clarity, it is important to note that humour theories and styles often serve as the foundational elements from which many models and frameworks are derived, and they also inform the development of scales used to measure humour. Figure \ref{Taxonomy} illustrates the dynamic nature of these aspects, showing how humour has been studied in relation to diverse contextual factors and research domains. This taxonomy not only organizes the existing body of knowledge but also highlights the multifaceted nature of humour, which is also the initial approach in applying these concepts to SE. The following subsections describe each key element of our preliminary taxonomy in detail. We link elements to SE team contexts to illustrate how they might be used to gain a better understanding of the impacts of humour in an SE context.

\subsection{Humour Theories}
Over the years, researchers from various disciplines including psychology, sociology, philosophy and linguistics have developed a range of theories to explain the mechanisms and effects of humour. We found a broad range of humour theories in the literature and categorise them into four groups based on their diverse domains and nature. \textcolor{black}{These theories  emphasise \textbf{why} something is perceived as humorous and describe the underlying psychological or cognitive mechanisms of humour.}

\subsubsection{Conventional Theories}
Among the diverse range of theories related to humour, \emph{conventional theories} have long been foundational in humour research and are often cited as the primary theories through which humour has been traditionally understood. 

\begin{itemize}
    \item \textbf{Incongruity Theory: } this is one of the foundational theories in humour research. \textcolor{black}{First introduced by Immanuel Kant \cite{kant2024critique, kant2008critique}}, this theory suggests that we find something humorous/funny when there is a mismatch between what we expect and what actually happens. This amusement we experience results in making people laugh. Incongruity theory was originated by the philosophers and explained that humour comes from the sudden shift when something we expect turns out be different \cite{kant2008critique, arthur2016world}. 
It is considered as an important theory because it helps researchers to understand how people minds process and react to things that are out of the ordinary or surprising, which is a big part of what makes something funny. However,  researchers also highlight that it does not pay attention to the other surrounding factors that might arise humour and why not all incongruities are funny \cite{mulder2002humour}. In an SE context, a design for a user interface that is incongruous may be found funny by the team, aid moreover aid them in improving it. \textcolor{black}{Further, this theory can be observed during code reviews when unexpected errors generate humour that alleviates tension and fosters collaboration}.
    
\end{itemize}

\begin{itemize}
    \item \textbf{Superiority Theory: } this is considered as one of the oldest concepts in humour research. \textcolor{black}{This theory is standardly attributed to Plato, Aristotle, and Thomas Hobbes, and later expanded by Gruner \cite{gruner2017game} }. The concept of superiority theory lies on the idea of explaining that people find something humorous/funny when it makes them feel superior to others. As per this theory, laughter often comes from the sense of dominance over others' mistakes or flaws. This has also been first originated by the philosophers and reformed which explains that every humorous situation has a winner or loser and humour requires an element of surprise. This is developed upon assumptions based on human nature and considered as significant because it sheds light on why certain types of humour, like teasing or sarcasm, are so common. It helps to understand how humour can be used not just for fun, but also as a way to establish social hierarchies or assert power over others \cite{gruner2017game}. In an SE context, this might be where a team expects certain information from a client but something unexpectedly humorous is found. However, care is needed as laughing at their client may be a negative rather than positive outcome.
\end{itemize}

\begin{itemize}
    \item \textbf{Relief Theory: } is developed based on nature of humour that serves as a mechanism to release pent-up tension or psychological energy. \textcolor{black}{First introduced by Spencer \cite{spencer1875physiology}, this theory gained prominence through Freud's work who proposed that humour and laughter allow for the spontaneous release of energy that accumulates due to the suppression of emotions or thoughts \cite{freud1960jokes}.} According to  studies, it not only provides physical relief but also offers psychological comfort by bringing repressed ideas to the surface in a socially acceptable way. The {relief theory} is often associated with the idea that laughter is beneficial to health as it alleviates the stress and tension that build up over time. It mainly focuses on these physiological and psychological impacts rather than explaining why certain things are perceived as humorous/ funny. Hence, researchers consider this more of a theory of laughter than humour \cite{mulder2002humour, morreall2011comic}. An SE team under undue time pressure may find relief by laughing at unachievable expectations. This may help improve individual and team welfare, as well as help the team dynamics. \textcolor{black}{For example, in high-pressure agile sprint retrospectives, humorous reflections on past challenges can release tension, helping teams discuss issues more openly and constructively.}
\end{itemize}

\subsubsection{Sociology Theories}

Sociological theories of humour explore how humour functions within social contexts, focusing on its role in shaping and reinforcing social structures. The humour research in sociology is not centered around the question why people laugh and view humour as more than just a source of entertainment. They see it as a powerful tool for social interaction, communication, and the maintenance of social norms and power dynamics.  
By examining humour through the lens of sociology, researchers can better understand how humour influences group dynamics, social hierarchies, and the ways in which individuals navigate social environments \cite{mulder2002humour}. This sociological humour has been categorised in three types of theories as follows;

\begin{itemize}
    \item \textbf{Maintenance Theory: } originates from the idea that humour serves to reinforce and sustain existing social structures and roles within a society. It suggests that jokes and humorous exchanges are crucial in maintain the status, particularly where there are clear social divisions such as workplace hierarchies, family structures and etc. In this theory, humour helps strengthen bonds within a group while also marking and reinforcing the differences between that group and others. This theory is important because it explains how humour can quietly reinforce social norms and biases, making it key to understanding the role of humour in maintaining social order \cite{davies1990ethnic, kuipers2008sociology}. In an SE team context, the database team might make fun at the expense of the UX team, or team members may joke about managers. While within the sub-team this might build rapport and perhaps help them counter poor management practices, for the overall team dynamics and organisational effectiveness it is likely to be counter-productive.
\end{itemize}

\begin{itemize}
    \item \textbf{Negotiation Theory: } views humour as a dynamic tool used in social interactions to navigate and negotiate relationships, statuses, and meanings. It suggests that humour can be employed to ease tensions, challenge authority, or subtly criticize social norms in a manner that is socially acceptable. Through playful exchanges, individuals negotiate their roles and identities, as well as the boundaries of acceptable behavior within a group. For instance, it is related to how stand-up comedians use humour to negotiate social issues and personal experiences, engaging audiences in a dialogue that can challenge or affirm societal norms \cite{gilbert2004performing}. This theory is significant because it underscores humour's capacity to facilitate social interaction and influence social change through nuanced and flexible communication \cite{kuipers2008sociology}. For example, SE team members might share amusing anecdotes about previous workplaces, and this may help with team bonding.
\end{itemize}

\begin{itemize}
    \item \textbf{Frame Theory: } suggests that humour involves shifting between different interpretive frames or perspectives, allowing people to view situations in a non-serious or playful context. This shift creates a psychological safe space where individuals can express criticism, or cope with stressful situations without facing direct confrontations. This has been introduced and developed by how people structure their experiences through various frames, including those that enable humour and irony. It is important because it explains how humour can temporarily alter our perception of reality, providing relief, insight, and opportunities for social commentary \cite{persson2018framing}. In a stressful SE team context, team members might create humorous anecdotes about current work pressures such as deadlines to ease tension. This might help in aiding individual and team morale and welfare, but may have negative impacts on productivity and overall organisational culture if it becomes cynical and overly negative.
\end{itemize}

\subsubsection{Psychology Theories}
Psychological theories of humour explore the mental and emotional processes that contribute to how we experience and understand humour. These theories focus on how our brains perceive, interpret, and respond to humorous stimuli. Unlike sociological theories, which emphasize the social context of humour, psychological theories delve into the cognitive and emotional mechanisms that make something funny to an individual. The study of humour from a psychological perspective helps us understand why certain types of humour resonate with people and how humour can affect our mental and emotional states. Two such psychology theories related to humour are as follows;

\begin{itemize}
    \item \textbf{Cognitive-Perceptual Theory: } suggests that humour arises from our ability to perceive and resolve incongruities—situations where there is a mismatch between our expectations and reality. Developed by Jerry Suls and further expanded by Rod A. Martin, this theory posits that humour occurs when we detect something unexpected or out of place and can mentally reconcile this incongruity \cite{martin2018psychology, suls1972two}. For instance, a joke often involves a setup that leads us to expect one outcome, but the punchline delivers something entirely different, creating a humorous effect. This theory is important because it highlights the role of cognitive processing in humour, explaining how our minds work to find humour in unexpected situations or wordplay.  In an SE work context, unexpected test outcomes might be found humorous if suggesting incongruous expectations.
\end{itemize}

\begin{itemize}
    \item \textbf{Emotional Response Theory: } focuses on how humour triggers various emotional reactions, which can range from joy and amusement to discomfort or embarrassment, depending on the type of humour and the context. Researchers have conducted studies on this theory examining the emotional impact of humour and how it can influence our feelings and moods. \cite{keltner1997study, ekman1992argument}. For example, a lighthearted joke might bring about feelings of joy and relaxation, while a sarcastic remark could cause discomfort or even offense. Emotional Response Theory is crucial for understanding the affecting side of humour—how it can be used to improve emotional well-being or, conversely, how it can lead to negative emotional outcomes if misused. SE teammates laughing at mistakes made by a colleague could be an example of negative outcomes for the target individual.
\end{itemize}

\subsubsection{Other Theories}

In addition to the conventional, sociological and psychological theories, there are some other theories that have emerged to explain diverse aspects of humour, often focusing on its playful, creative or unpredictable nature. Following are some prominent theories which provide insights into how humour functions in various contexts. These are crucial in understanding how humour can serve as a tool for dealing with potentially sensitive or controversial topics.

\begin{itemize}
    \item \textbf{Play Theory: } was primarily developed by Johan Huizinga and later expanded by Brian Sutton-Smith, suggests that humour is a form of play that allows individuals to experiment with social boundaries and engage in creative thinking \cite{huizinga1955study, sutton2001ambiguity}. According to this theory, humour creates a space where the usual rules and norms are temporarily suspended, enabling people to explore new ideas and perspectives without the usual constraints of reality. This playful nature of humour is evident in creative industries where humour is often used to inspire innovation and in educational settings where it can facilitate learning by making complex or abstract concepts more accessible \cite{ziv1988teaching, meyer2000humor}. Play Theory is significant because it highlights the role of humour in fostering creativity and adaptability in both social and cognitive processes. In UI design work, SE team members playing with unusual design ideas could enhance their design outcomes.
\end{itemize}

\begin{itemize}
    \item \textbf{Benign Violation Theory: } introduced by Peter McGraw and Caleb Warren, suggests that humour arises when a situation is perceived as both a violation of some norm or expectation and simultaneously benign or non-threatening \cite{mcgraw2010benign}. This theory helps explain why people can find humour in situations that involve controversial topics, as long as these situations are framed in a way that makes them feel safe or acceptable. For example, a joke about a sensitive subject might be funny if it is delivered in a way that softens the potential offense, making it a "benign" violation. This theory is important because it helps understand how humour can navigate the fine line between what is acceptable and what is offensive, and why some jokes succeed while others fail. In an SE team, this might be used to pass on critical feedback in a `safe' manner.

\end{itemize}

\begin{itemize}
    \item \textbf{Minsky's Theory: } Marvin Minsky's theory of humour is often associated with cognitive science. This theory suggests that humour arises when the brain detects a contradiction between two mental schema or frames of reference. According to Minsky, humour occurs when the mind unexpectedly shifts from one interpretation of a situation to another, revealing an incongruity that was previously unnoticed. This sudden shift and the realization of the incongruity produce the humorous effect. Minsky's theory is significant because it connects humour to the broader workings of the mind, particularly in how we process and resolve contradictions in our perceptions of the world. This theory is could be relevant in fields like AI software development, where understanding human humour can inform the development of more sophisticated and human-like machines \cite{minsky1988society, mulder2002humour}.
\end{itemize}

\subsection{Humour Styles}

Humour styles refer \textcolor{black}{to \textbf{how} humour is expressed in interpersonal contexts and its impact on social dynamics. These styles emphasis
 various ways individuals express and utilize humour, which can significantly influence social interactions, psychological well-being, and personal relationships.} These styles are broadly categorized into positive, negative, and contextual types, each with distinct impacts on individuals and groups. 
Research into these humour styles, especially through tools like the humour Styles Questionnaire (HSQ), \textcolor{black}{developed by Martin et al. \cite{martin2003individual} has provided valuable insights into the complex relationships between humour, mental health, social bonds, and personality traits.} The following sub-sections explore these humour styles.

\subsubsection{Positive Humour}
Positive humour includes styles like affiliative and self-enhancing humour, which are used to promote social bonding and maintain a positive outlook in challenging situations. These forms of humour are generally associated with improved mental health, stronger social connections, and greater resilience.

\begin{itemize}
    \item \textbf{Affilliative Humour: }  refers to humour that has been used to enhance social interactions and relationships. This style involves making lighthearted jokes and sharing funny anecdotes to promote social cohesion, reduce tensions, and create a positive group atmosphere \cite{martin2003individual}. Research has shown that affiliative humour is positively correlated with psychological well-being and is effective in fostering close interpersonal relationships \cite{dyck2013understanding}.  Studies indicate that individuals who frequently use affiliative humour tend to be more extroverted, sociable, and have a greater sense of belonging within their social groups \cite{kuiper2004humor}. The humour Styles Questionnaire (HSQ), developed by Rod Martin et al. \cite{martin2003individual}, is one of the key instruments used to measure affiliative humour and its impact on social relationships and mental health. In an SE team it may be used to create a sense of camaraderie and to defuse potential conflicts within the team \cite{vernon2008behavioral}. \textcolor{black}{For example, in daily stand-ups, sharing lighthearted anecdotes about development blockers (e.g., “My code went on vacation—still not returning results!”) can foster a supportive environment. Further, SE teams frequently use memes or humorous Slack reactions to promote a positive team culture and encourage engagement during remote collaboration.} 
\end{itemize}

\begin{itemize}
    \item \textbf{Self-enhancing Humour: }  involves maintaining a humorous outlook on life, particularly in the face of adversity. This style is characterized by a tendency to use humour to cope with stress and to maintain a positive perspective \cite{martin2003individual}. Research suggests that self-enhancing humour is associated with higher levels of emotional well-being and resilience, as individuals using this style are better equipped to handle life's challenges without becoming overwhelmed. It has also been linked to lower levels of anxiety and depression, making it a valuable tool for psychological resilience \cite{dyck2013understanding, kuiper2004humor}. This could benefit individuals and team members in SE to e.g. to better handle stressful requirements change requests \cite{madampe2022emotional}, and may help to enhance team dynamics and morale.
\end{itemize}

\subsubsection{Negative Humour}
Negative humour encompasses styles such as aggressive and self-defeating humour, which can create tension or harm relationships by targeting others or oneself. While sometimes used to cope with stress, these types of humour are often linked to lower self-esteem, increased conflict, and negative psychological outcomes .

\begin{itemize}
    \item \textbf{Aggressive Humour: } involves using humour to criticize or manipulate others, often in a way that can be hurtful or belittling. This style includes sarcasm, teasing, and ridicule, and is often used to assert dominance \cite{martin2003individual}. While some studies suggest that aggressive humour can sometimes enhance group dynamics by reinforcing social norms or relieving tension, it is generally associated with negative social outcomes. Research has linked aggressive humour with higher levels of interpersonal conflict, lower levels of social support, and increased incidences of bullying and workplace incivility \cite{dyck2013understanding, kuiper2004humor, vernon2008behavioral}. This may manifest in an SE team as sarcastic humour from one individual to another, negatively impacting team dynamics and individual well-being.
\end{itemize}

\begin{itemize}
    \item \textbf{Self-defeating Humour: } is characterized by self-disparaging jokes or comments made at one's own expense, often to gain social approval or to fit in with a group. While it can sometimes serve to defuse tension or to ingratiate oneself with others, it is generally linked to negative psychological outcomes \cite{martin2003individual}. Studies have found that self-defeating humour is associated with lower self-esteem, higher levels of depression, and a greater propensity for anxiety. It is often used by individuals who feel insecure or who are seeking to avoid confrontation, and can perpetuate negative self-perceptions and reinforce harmful social dynamics \cite{dyck2013understanding, vernon2008behavioral}. An SE team member making fun of or belittling their own work or work effort is an example of such self-defeating humour.
\end{itemize}

\subsubsection{Contextual Humour}
Contextual humour, including situational and dark humour, is heavily influenced by the specific environment and circumstances in which it is used. These styles can either enhance group cohesion in shared contexts or, if misused, lead to discomfort or misunderstanding, especially when dealing with sensitive topics .

\begin{itemize}
    \item \textbf{Situational Humour: } is a type of humour that arises from specific circumstances or contexts, often relying on the peculiarities of a particular situation to generate amusement. This form of humour is highly dependent on timing and the environment, making it particularly effective in social settings where participants share a common understanding of the context \cite{apte1985humor}. Studies have shown that situational humour can enhance group cohesion by providing a shared experience that reinforces group identity and solidarity. However, because it is context-specific, situational humour can sometimes lead to misunderstandings or feelings of exclusion among those who are not familiar with the particular situation being referenced \cite{mulkay1988humour}. SE team members making fun of software artefacts might be an example of positive situation humour, but doing this in the presence of non-SE stakeholders be an example of the later.
\end{itemize}

\begin{itemize}
    \item \textbf{Dark Humour: } also known as black humour, involves making light of topics that are generally considered serious, or painful, such as death, war, or illness. This type of humour is often used as a coping mechanism, allowing individuals to confront uncomfortable truths in a less threatening way. Research has shown that dark humour can serve as a means of psychological defense, helping individuals to process and make sense of difficult experiences. However, dark humour can also be polarizing, as it may be perceived as offensive or inappropriate by those who do not share the same perspective or cultural background \cite{lefcourt2001humor}. SE team members working on digital health applications for serious diseases making fun of disease outcomes might be such an example. While this could have some benefit in terms of coping in the face of serious health issues, it may result in negative team performance outcomes.
\end{itemize}

\subsection{Humour Models and Frameworks}
Humour models and frameworks are conceptual tools used to understand and analyze the various aspects of humour, including its origins, effects, and applications. These models are designed to provide a structured way to study humour in different contexts, such as psychology, sociology, education, and organizational behavior.  


\begin{itemize}
    \item \textbf{Wheel Model of Humour: } is a cyclical framework that describes how humour events lead to positive emotional responses, which in turn reinforce a humour-supportive environment. This model emphasizes the feedback loop created by humour, where the use of humour can foster an atmosphere that encourages further humour, enhancing group cohesion and emotional well-being \cite{robert2012wheel}. Aaker and Bagdonas have utilized this model to analyze humour in organizational settings, demonstrating how humour can improve team dynamics and leadership effectiveness by promoting a positive and inclusive workplace culture \cite{aaker2021humor}. Similarly, this model may assist with enhancing SE team dynamics.
\end{itemize}

\begin{itemize}
    \item \textbf{Group Humour Effectiveness Model (GHEM): } explains how humour can positively impact group effectiveness by enhancing communication, leadership, and emotional management. This model is particularly relevant in team-based work environments, where humour can serve as a tool to reduce tension, build trust, and improve collaboration. Researchers have explored the application of GHEM in leadership studies, highlighting how humour can be used strategically to navigate complex social dynamics and foster a more cohesive and productive team environment \cite{romero2008humor}. This model seems well suited to SE team management and team climate development.
\end{itemize}

\begin{itemize}
    \item \textbf{Pedagogical Humour Model: } explains how humour can be used as an educational tool to enhance learning and student engagement. This model is particularly relevant in classroom settings, where humour can be strategically employed to make learning more enjoyable and effective. Researchers have explored how humour can improve student retention, motivation, and understanding of complex concepts, particularly in higher education \cite{berk2003professors}. SE education may benefit from use of the model, particularly when introducing challenging SE topics.
\end{itemize}

\begin{itemize}
    \item \textbf{Humour Styles Framework: } classifies humour into different styles based on how it is used and its impact on social relationships. This framework is used for analyzing the role of humour in personal interactions and personality assessment, as it differentiates between positive and negative humour styles, such as affiliative, self-enhancing, aggressive, and self-defeating humour. Researchers have conducted extensive research on this framework, showing how different humour styles can influence mental health, social bonds, and overall well-being \cite{martin2003individual}.
\end{itemize}

\subsection{Humour Scales}

Humour scales are tools used to measure an individual's sense of humour, humour styles, and the impact of humour on various psychological and social outcomes. These scales are developed based on humour theories, styles, models or frameworks and often utilized in psychological assessments and research to understand how humour influences well-being, coping mechanisms, and social interactions. We have grouped these scales based on how they measure humour, namely, self-report measures, multi-dimensional measures, and trained observer measures. Each group offers different approaches to assessing humour, providing a comprehensive understanding of how humour functions in diverse contexts. 

\subsubsection{Self-Report Humour Measures}

\begin{itemize}
    \item \textbf{Sense of Humour Questionnaire: } is a self-report measure designed to assess an individual's overall sense of humour. This scale evaluates how frequently a person engages in humour, their attitudes toward humour, and how they perceive humour in social situations. In its commonly used form, the SHQ consists of 7 items that measure three primary aspects of humour: humour production (how often a person uses humour), humour perception (how often a person finds things funny), and attitude toward humour (how positively a person views humour in general). Participants respond to each item on a Likert scale, indicating the frequency or extent to which they engage in or experience the aspects of humour being assessed. The responses are then scored to provide an overall sense of humour score, with higher scores indicating a greater sense of humour. The SHQ is widely used in research to explore the relationship between humour and various psychological outcomes, such as stress resilience, social competence, and mental health \cite{svebak1996development}.
\end{itemize}

\begin{itemize}
    \item \textbf{Coping Humour Scale (CHS): } assesses the extent to which individuals use humour as a coping mechanism in stressful situations. This scale is particularly useful in studies examining the role of humour in stress management and psychological well-being. 
    The scale consists of 7 items that participants rate on a 4-point Likert scale, ranging from "strongly disagree" to "strongly agree." These items assess how frequently individuals use humour to re-frame stressful situations in a positive light \cite{martin1983sense}.
The CHS has been widely utilized in studies investigating the impact of humour on mental health, particularly in relation to stress-related outcomes such as anxiety and depression. Research has consistently shown that individuals with higher CHS scores tend to experience better psychological well-being, as humour enables them to cope more effectively with stress by providing a means to distance themselves from the emotional intensity of challenging situations. The CHS has proven to be a reliable predictor of positive health outcomes, making it an essential tool in the field of psychological research on coping mechanisms. \cite{pandin2023influence, abel1998interaction}. SE team members working in stressful situations may benefit from showing high CHS.

\end{itemize}

\begin{itemize}
    \item \textbf{Situational Humour Response Questionnaire (SHRQ): } is designed to measure how frequently individuals experience humour in response to everyday situations. Unlike other humour scales that primarily focus on humour production, the SHRQ emphasizes the reception and perception of humour, making it unique in its approach to understanding humour's role in daily life. This includes 18 items, each a different everyday situation, and respondents are asked to rate how often they would find each situation humorous on a 5-point Likert scale ranging from 1 (never) to 5 (very often). The situations presented in the SHRQ cover a wide range of social and personal contexts, such as interactions with friends, encountering unexpected events, or dealing with minor inconveniences. For example, items might include scenarios like "You accidentally spill something on yourself in public" or "A friend tells a joke that others find funny." The aim of the SHRQ is to assess how likely an individual is to perceive humour in these types of situations, rather than their ability to produce humour. This scale has been particularly useful in research exploring the relationship between humour perception and psychological health, including studies on stress resilience and social well-being \cite{martin1996situational}. 

\end{itemize}

\begin{itemize}
    \item \textbf{Multidimensional Sense of Humour Scale (MSHS): } is designed to assess various dimensions of humour, including humour appreciation, humour creation, and the use of humour as a coping mechanism. The MSHS consists of 24 items divided into four sub-scales: humour appreciation, humour production, coping humour, and attitude toward humour. Each item is rated on a 5-point Likert scale, where respondents indicate their agreement with statements like "I appreciate jokes even if I have heard them before" or "I often find myself making others laugh." The MSHS is particularly valuable in research that explores the relationship between humour and mental health, as it provides a comprehensive understanding of how humour manifests across different aspects of life \cite{thorsonmultidimensional}.
\end{itemize}

\subsubsection{Multi-dimensional Measures}

\begin{itemize}
    \item \textbf{Humour Style Questionnaire (HSQ): } is the most widely used multi-dimensional measure that classifies humour into four distinct styles: affiliative humour, self-enhancing humour, aggressive humour, and self-defeating humour. The HSQ is developed based on humour style framework which consists of 32 items, with 8 items corresponding to each humour style. Participants rate each item on a 7-point Likert scale (ranging from 1 = totally disagree to 7 = totally agree). Example items include statements like "I enjoy making people laugh" (affiliative humour), "If I am feeling depressed, I can usually cheer myself up with humour" (self-enhancing humour), "If someone makes a mistake, I will often tease them about it" (aggressive humour), and "I often put myself down when I try to be funny" (self-defeating humour). The HSQ is extensively used in studies that explore the role of humour in mental health, social functioning, and personality traits \cite{martin2003individual}.
\end{itemize}

\subsubsection{Trained Observer Measures}

\begin{itemize}
    \item \textbf{Humorous Behaviours Q-Sort Deck: } is a trained observer measure which is used to assess humour behaviors in social interactions. This tool involves trained raters who observe and categorize humour-related behaviors exhibited by individuals in various settings. The Q-Sort Deck includes 100 statements describing different humour behaviors, such as "Makes witty remarks during conversations" or "Uses humour to defuse tense situations." Raters sort these behaviors based on how frequently they are observed, creating a profile of the individual's humour style. This method provides an objective assessment of humour in naturalistic settings and is particularly useful in studies examining humour in group dynamics, leadership, and communication \cite{ruch1998two}.
\end{itemize}

\subsection{Contextual Factors in Humour Research}

Humour research is profoundly influenced by contextual factors, which shape how humour is perceived, applied, and interpreted across different environments. 
Some of the most prominent factors are \textbf{cultural differences}. Humour varies widely across cultures, with what might be considered amusing in one culture potentially being offensive or incomprehensible in another. This diversity in culture creates challenges in cross-cultural communication and global team dynamics, making it crucial for researchers and practitioners to understand the nuances of humour within different cultural contexts. For instance, studies have been conducted on these cultural variances, highlighting the importance of context in understanding humour's impact on communication and interpersonal relationships \cite{davies1990ethnic}.

 \textbf{Industry-specific applications} of humour highlight that different industries have varying tolerances and uses for humour, reflecting the norms and expectations within each sector. For example, humour is often more freely accepted and utilized in creative industries such as advertising or entertainment, where it can be a tool for innovation and engagement. Conversely, in more conservative fields, like finance or sales, humour may be less prevalent or more restrained due to professional norms and the serious nature of the work. Studies have explored how humour operates within these industry-specific contexts, offering insights into how organizational culture and professional conduct shape the use of humour in the workplace \cite{bergeron2008effects, pham2021laughing, al2012contextual}. The operation of humour in software engineering in general, software engineering teams in particular domains, and software engineering for different stakeholders, could all be very insightful.

\textbf{Team dynamics} play a crucial role in how humour is used and received in organizational settings. The composition of a team—including factors such as size, hierarchy, and diversity—can influence the effectiveness of humour as a tool for enhancing group cohesion, leadership, and conflict resolution. humour can serve as a powerful mechanism for fostering a positive team environment, but it must be carefully managed to ensure it is inclusive and does not alienate team members. Research has shown that humour, when used appropriately, can strengthen team dynamics and improve overall group performance \cite{romero2008humor}. In virtual teams, where non-verbal cues are often absent, humour can bridge communication gaps but also runs the risk of being misinterpreted, highlighting the need for careful consideration of how humour is employed in remote work environments \cite{lehmann2014fun}.

\textcolor{black}{Although \textbf{age and gender} differences are also essential contextual factors, they are often intertwined with cultural norms and team dynamics, shaping humour preferences and sensitivities. For example, generational differences in humour style may emerge during team interactions, and gender norms can influence the acceptability of certain types of humour. While we do not expand on them separately in this section, we acknowledge their importance and encourage further SE-specific studies on their impact on team humour dynamics}.

\section{Discussion} \label{discussion}
\textcolor{black}{We developed a preliminary taxonomy of humour based on the humour literature. We identified that although there are diverse set of humour theories, styles, models, and scales that have been used in diverse contexts, there is no unified framework for humour that discuss the potential use of humour in the SE context. For example, studies in psychology and organizational behaviour have developed models such as the Group Humour Effectiveness Model (GHEM) and the Wheel Model of Humour, focusing on the impact of humour on group dynamics. However, these models are conceptual frameworks rather than comprehensive taxonomies. Frameworks such as the Pedagogical Humour Model emphasize humour’s role in learning outcomes, but these frameworks are context-specific and not structured as taxonomies. Our taxonomy is novel because it integrates four key elements—theories, styles, models, and scales—into a structured framework, unifying cross-disciplinary insights into a single framework applicable to SE practices and provides a foundation for future empirical testing of humour’s role in SE. This structured approach is helpful in understanding humour's dynamic nature across various SE contexts, SE team members, and SE stakeholders. }Traditionally, SE is perceived as a highly technical and serious discipline, but this often overlooks the human and social dynamics that play a crucial role in team collaboration and project success. Our aim is to address this gap by providing SE researchers and practitioners with a preliminary framework to explore how the presence of humour can be strategically employed to enhance team dynamics, improve communication, and foster a positive organizational culture.

\textcolor{black}{As we discussed in section \ref{related work} , a small number of the reviewed papers directly examine humour within SE contexts. For example, Khan et al. \cite{khan2022leading} apply humour styles (affiliative, self-enhancing, aggressive, and self-defeating) to assess their impact on creativity in software teams and use the HSQ scale developed by Martin et al. \cite{martin2003individual}. The study on humour and developer engagement \cite{tiwari2024great} discusses humour in tools like \emph{faker} and \emph{lolcommits}, implicitly reflecting relief and incongruity theories, though these are not explicitly named. Krienke and Bansal \cite{krienke2017information} observe humour during requirements elicitation interviews, where affiliative and self-enhancing humour styles are evident, but the work lacks theoretical framing or structured measurement. None of these studies apply formal models such as the Wheel Model or Group Humour Effectiveness Model. This shows that while humour appears in SE literature, its exploration remains partial, typically limited to styles or surface-level observations and highlights a lack of theory-based, model-informed, and measurement-driven approaches. The taxonomy thus provides a structured lens to build more coherent and transferable research in this emerging space.}

\textbf{Suitability and applicability of the Taxonomy in the SE context}:
Our taxonomy presents a broad based understanding of humour theories, styles, models, and frameworks, offering SE professionals a toolkit to better understand and leverage humour in their work environments. For instance, the application of conventional humour theories, such as incongruity and relief, within SE can help teams navigate complex problem-solving scenarios by introducing humour to diffuse tension and promote creative thinking. In practice, this could mean incorporating humour in brainstorming sessions to encourage innovative ideas or using humour during code reviews to ease the critique process, making it more constructive rather than confrontational. \textcolor{black}{For example, humour styles such as affiliative and self-enhancing could be encouraged in code reviews or retrospective meetings, where tensions can easily arise. Rather than neutralising feedback, humour in the form of light-hearted comments or shared jokes might promote psychological safety and reduce defensiveness. The Wheel Model of Humour could be used to understand how recurring of positive humour during stand-up meetings contributing to a humour-supportive team climate over time. In requirements elicitation, where misunderstandings between stakeholders and engineers are common, incongruity theory may explain why humour is used to re-frame problems or highlight mis-alignments in a non-confrontational way.}

\textcolor{black}{In the context of scrum ceremonies (e.g., daily stand-ups or sprint planning), humour may support group viability by building cohesion (as per the Group Humour Effectiveness Model). For instance, team leads using affiliative humour to engage quieter members may help surface unspoken blockers or risks. In bug reporting and testing, testers often use subtle humour (e.g., sarcastic bug titles, memes in issue trackers) as a means of reducing the emotional load of reporting recurring or trivial defects—this aligns with self-enhancing humour and may help with emotional regulation in repetitive QA tasks. Similarly, including humour in on-boarding materials or documentation (e.g., light jokes in internal wikis or tutorials) may lower cognitive barriers and make learning more accessible for junior developers.}

\textcolor{black}{These examples illustrate how humour in SE is not just social filler but can functionally support communication, feedback, learning, and emotional management. Since the taxonomy integrates theoretical and empirical elements, SE researchers could design studies such as aligning incongruity theory with affiliative humour styles during code reviews, and measure their impact on team cohesion using the HSQ scale. The taxonomy provides a structured basis to conduct such investigations. However, while the taxonomy provides a robust framework, its categories have not yet been empirically validated in the SE context. This raises questions about the applicability of humour theories, styles, models, and scales in SE environments, which may differ in communication norms and team structure from domains like psychology or education. Future research should focus on testing how humour impacts teamwork, productivity, creativity, and emotional resilience across different SE roles and practices. This would help evaluate the practical relevance of the taxonomy in diverse software settings.}

\textbf{Potential applications and future research in SE}:  
\textcolor{black}{To advance SE research, future studies should adopt a holistic approach using the taxonomy. For example, a mixed-method study could apply superiority theory to analyze humour in stand-up meetings, categorize humour styles through observations, and measure impact with HSQ. Longitudinal studies could further investigate how humour styles guided by relief theory alleviate burnout during project cycles.}
Given the potential benefits of humour, SE organizations might consider incorporating humour training and assessment into their professional development programs. Such training could involve educating team members on the different humour styles and how to apply them appropriately within the workplace. For example, training could focus on using humour to enhance communication during stand-up meetings or retrospectives, where sharing lighthearted moments can help team members bond and reduce the stress of tight deadlines. Research indicates that when humour is used appropriately, it can lead to higher job satisfaction, improved employee engagement, and better overall performance \cite{romero2008humor}. However, the effectiveness of these training programs in SE contexts has not yet been established, representing a significant opportunity for future research.
The taxonomy also highlights several untapped possibilities in evaluating the efficacy of utilizing humour to enhance SE-related activities. Future studies could explore how humour training might improve developers' ability to empathize with users or how humour might be used to make technical education more engaging for students.

\textbf{Responsible use of humour in SE}:
A responsible use of humour is crucial in maintaining a healthy and productive SE environment. Given the diversity in SE teams, humour that is culturally sensitive and inclusive can help bridge communication gaps and strengthen team dynamics. Our taxonomy acts as an initial step for understanding these nuances, emphasizing that humour should be used to build rather than divide teams. This is particularly relevant in global SE teams, where cultural differences can influence how humour is perceived and received \cite{davies1990ethnic}. Additionally, the taxonomy's inclusion of contextual humour, such as situational and dark humour, offers insights into how humour can be adapted to different scenarios within SE. For example, situational humour can be particularly effective in pair programming or during long coding sessions, where it can lighten the mood and maintain focus. However, SE professionals must be mindful of the boundaries of humour, ensuring it does not alienate or offend team members.

\textbf{Challenges and limitations}:
Each of the humour scales we included in our taxonomy, such as the Humour Styles Questionnaire (HSQ) and Coping humour Scale (CHS), has its strengths and weaknesses, making it challenging for researchers to determine the best-fitting scale for use in SE contexts. For example, while the HSQ has been applied in studies conducted in Pakistan to explore the role of humour in software organizations \cite{amjed2016effect}, its broader applicability in SE contexts remains uncertain. The generalizability of these scales to SE settings, particularly regarding how different humour styles impact team dynamics, creativity, and stress management, needs further exploration. Whether and how to generalize these models and what sort of studies should be conducted to evaluate their effectiveness in SE contexts are areas ripe for investigation.

Moreover, our humour taxonomy underscores the importance of finding the correct balance of humour in SE. While humour can be a powerful tool for enhancing team dynamics and reducing stress, it must be used responsibly to avoid potential pitfalls. Over-reliance on humour, particularly in formal settings like client meetings or critical decision-making processes, could undermine professionalism and lead to misunderstandings. The taxonomy's detailed breakdown of humour styles and their impact on social interactions provides valuable guidance for SE professionals to strike the right balance—using humour to enhance, rather than detract from, their work. Furthermore, in the SE context, where teams are often diverse and work under high-pressure conditions, humour should be used inclusively and respectfully.

\textcolor{black}{While our preliminary taxonomy of humour provides insights for SE researchers and practitioners, it also highlights the need for further validation and exploration in SE-specific contexts. Our taxonomy offers a valuable framework for understanding humour in SE, but several limitations stem from the scope and methodology of this study. The taxonomy was developed from existing literature across multiple disciplines, including psychology, organizational behavior, and education. This cross-disciplinary approach, while valuable, may have overlooked humour styles or models uniquely relevant to SE practices. SE teams often exhibit distinct cultural and professional norms that may influence humour patterns. Future research should address this gap by conducting SE-specific qualitative studies such as interviews, focus groups, or ethnographic observations to capture humour patterns unique to SE teams. Additionally, the taxonomy is based solely on prior literature without empirical testing within SE settings. This limits its immediate applicability to SE teams and raises the need for empirical validation. To strengthen its validity, future research should include:
\begin{itemize}
    \item Pilot studies to apply the taxonomy in SE teams and assess its relevance.
    \item Case studies to observe humour’s influence on SE collaboration and productivity.
    \item Controlled experiments using humour scales (e.g., HSQ and CHS) to evaluate humour’s impact on SE team performance.
\end{itemize}
Further, while our taxonomy categorizes established humour scales such as the Humour Styles Questionnaire (HSQ) and Coping Humour Scale (CHS), these scales were designed for general organizational contexts and may not fully capture humour’s nuances within SE teams. For instance, humour styles that foster creativity in pair programming may not have the same impact during sprint retrospectives. Therefore, we suggest that future research should develop SE-specific humour scales or adapt existing scales to SE practices, conduct comparative studies across different professional contexts to assess differences in humour application and perform longitudinal studies to understand how humour evolves within SE teams over time.
Moreover, while the taxonomy offers a starting point for integrating humour into SE practices, its full potential can only be realized through empirical studies that examine its applicability and effectiveness within SE practice. Future research should focus on testing the taxonomy’s components in real-world SE settings, exploring how humour can be used to enhance various aspects of SE practice, from team dynamics to stress management.}

\textcolor{black}{Finally, it is essential to develop and validate humour scales specifically tailored to SE contexts, ensuring that researchers and practitioners have the tools they need to measure and apply humour effectively in their work. Such scales should be capable of capturing humour’s impact across different SE activities (e.g., stand-up meetings, pair programming, and retrospectives) and be adaptable to diverse SE methodologies such as Agile, DevOps, and Waterfall. In summary, our taxonomy highlights both opportunities and challenges in integrating humour into SE practice. Addressing these limitations through empirical studies, SE-specific scales, and cross-contextual comparisons will be crucial for advancing both research and practice in this emerging area.}

\section{Conclusion} \label{conclusion}

This paper introduces a literature review-based preliminary taxonomy of humour as it may be applied within the SE context. Recognizing that SE has traditionally been viewed as a highly technical and serious discipline, our taxonomy aims to bring attention to the to date very overlooked human and social humour dynamics that  may significantly influence team collaboration and project outcomes. By categorizing humour into distinct theories, styles, models, and scales, our taxonomy provides SE researchers and practitioners with a structured framework to explore how humour can be strategically employed to enhance various aspects of software development, including team dynamics, communication, and stress management. Our taxonomy is grounded in extensive literature from diverse fields such as psychology, sociology, and organizational behavior, but its direct applicability to SE is yet to be fully validated. While our study highlights the potential benefits of humour in SE, it also acknowledges the challenges of generalizing these concepts to a field where empirical evidence is still limited. The categories presented, such as humour styles and models, offer valuable insights, but further research is needed to evaluate their effectiveness in SE-specific contexts. Moreover, the taxonomy underscores the importance of a balanced and responsible use of humour in SE. While humour can be a powerful tool for fostering a positive work environment, it must be applied thoughtfully to avoid potential pitfalls, such as misunderstandings or the alienation of team members. The inclusion of humour scales, like the humour styles questionnaire (HSQ) points to the need for more targeted research within SE to better understand how these scales can be used to measure and enhance humour's impact on team performance and individual well-being.
We encourage SE professionals and researchers to investigate the role of humour more deeply. We hope that our humour taxonomy will contribute to the development of more cohesive, creative, and psychologically supportive SE environments.

\section*{Acknowledgments}
Hidellaarachchi and Grundy are supported by ARC Laureate Fellowship FL190100035

\bibliographystyle{elsarticle-num} 
\bibliography{output}

\section*{Appendix A: Papers List} 

\begin{table*}[]
\centering
\caption{\centering List of papers reviewed }
\label{TABLE: paper list}
\resizebox{\textwidth}{!}{%
\begin{tabular}{@{}ll@{}}
\toprule
\multicolumn{1}{l}{\textbf{\begin{tabular}[c]{@{}l@{}}Paper ID\end{tabular} } }      & \textbf{\begin{tabular}[c]{@{}l@{}} Paper\end{tabular}}                            \\ \midrule

[1] & \begin{tabular}[c]{@{}l@{}} E. Romero, A. Pescosolido, Humor and group effectiveness, Human relations 61 (3) (2008) 395–418 (\cite{romero2008humor})
 \end{tabular} \\  

[2] & \begin{tabular}[c]{@{}l@{}} C. Robert, J. E. Wilbanks, The wheel model of humor: Humor events and affect in organizations, \\Human Relations 65 (9) (2012) 1071–1099 (\cite{robert2012wheel})\end{tabular} \\

[3] & \begin{tabular}[c]{@{}l@{}} B. Plester, Crossing the line: Boundaries of workplace humour and fun, Employee Relations 31 (6) (2009) 584–599.(\cite{plester2009crossing}) \end{tabular} \\

[4] & \begin{tabular}[c]{@{}l@{}} S. Moussawi, R. Benbunan-Fich, The effect of voice and humour on users’ perceptions of personal \\intelligent agents, Behaviour \& Information
Technology 40 (15) (2021) 1603–1626.(\cite{moussawi2021effect})\end{tabular} \\

[5] & \begin{tabular}[c]{@{}l@{}} C. Watson, V. Drew, Humour and laughter in meetings: Influence, decision-making and the emergence \\of leadership, Discourse \&
Communication 11 (3) (2017) 314–329.(\cite{watson2017humour})\end{tabular} \\

[6] & \begin{tabular}[c]{@{}l@{}} B. Plester, Healthy humour: Using humour to cope at work, Kotuitui: New Zealand Journal of Social \\Sciences Online 4 (1) (2009) 89–102.(\cite{plester2009healthy}) \end{tabular} \\

[7] & \begin{tabular}[c]{@{}l@{}} M. H. Abel, Interaction of humor and gender in moderating relationships between stress and outcomes, \\The Journal of psychology 132 (3)
(1998) 267–276. (\cite{abel1998interaction})\end{tabular} \\

[8] & \begin{tabular}[c]{@{}l@{}} C. Robert, W. Yan, The case for developing new research on humor and culture in organizations: Toward a \\higher grade of manure, in:
Research in personnel and human resources management, 2007, pp. 205–267.(\cite{robert2007case})\end{tabular} \\

[9] & \begin{tabular}[c]{@{}l@{}} A. Cann, C. L. Zapata, H. B. Davis, Positive and negative styles of humor in communication: Evidence for\\ the importance of considering
both styles, Communication Quarterly 57 (4) (2009) 452–468(\cite{cann2009positive})\end{tabular} \\

[10] & \begin{tabular}[c]{@{}l@{}} J. Bergeron, M.-A. Vachon, The effects of humour usage by financial advisors in sales encounters,\\ International Journal of Bank Marketing
26 (6) (2008) 376–398.(\cite{bergeron2008effects}) \end{tabular} \\

[11] & \begin{tabular}[c]{@{}l@{}} B. Bitterly, A. W. Brooks, Sarcasm, self-deprecation, and inside jokes: A user’s guide to humor at work,\\ Harvard Business Review 98 (4)
(2020) 96–103.(\cite{bitterly2020sarcasm}) \end{tabular} \\

[12] & \begin{tabular}[c]{@{}l@{}} R. A. Martin, P. Puhlik-Doris, G. Larsen, J. Gray, K. Weir, Individual differences in uses of humor and their \\relation to psychological
well-being: Development of the humor styles questionnaire, Journal of research \\in personality 37 (1) (2003) 48–75.(\cite{martin2003individual})\end{tabular} \\

[13] & \begin{tabular}[c]{@{}l@{}} M. P. Mulder, A. Nijholt, Humour research: State of art (2002)(\cite{mulder2002humour})\end{tabular} \\

[14] & \begin{tabular}[c]{@{}l@{}} N. Lehmann-Willenbrock, J. A. Allen, How fun are your meetings? investigating the relationship between \\humor patterns in team interactions
and team performance., Journal of Applied Psychology 99 \\(6) (2014) 1278.(\cite{lehmann2014fun})\end{tabular} \\

[15] & \begin{tabular}[c]{@{}l@{}} M. Hoffmann, D. Mendez, F. Fagerholm, A. Luckhardt, The human side of software engineering teams:\\ an investigation of contemporary
challenges, IEEE Transactions on software engineering 49 (1) \\(2022) 211–225.(\cite{hoffmann2022human}) \end{tabular} \\

[16] & \begin{tabular}[c]{@{}l@{}} V. Borsotti, P. Bjørn, Humor and stereotypes in computing: An equity-focused approach to institutional \\accountability, Computer Supported
Cooperative Work (CSCW) 31 (4) (2022) 771–803(\cite{borsotti2022humor})\end{tabular} \\

[17] & \begin{tabular}[c]{@{}l@{}} D. Krienke, G. Bansal, Information system requirement elicitation: The role of humor (2017)(\cite{krienke2017information})\end{tabular} \\

[18] & \begin{tabular}[c]{@{}l@{}} C. Christman, Instructor humor as a tool to increase student engagement (2018)(\cite{christman2018instructor})\end{tabular} \\

[19] & \begin{tabular}[c]{@{}l@{}}  D. M. R. Barros, L. R. Begosso, J. A. Fabri, A. L’Erario, The use of comic strips in the teaching \\of software engineering, in: 2017 IEEE
Frontiers in Education Conference (FIE), IEEE, 2017, pp. 1–8.(\cite{barros2017use})\end{tabular} \\

[20] & \begin{tabular}[c]{@{}l@{}} D. Tiwari, T. Toady, M. Monperrus, B. Baudry, With great humor comes great developer engagement, \\in: Proceedings of the 46th International
Conference on Software Engineering: Software Engineering in Society,\\ 2024, pp. 1–11.(\cite{tiwari2024great})\end{tabular} \\

[21] & \begin{tabular}[c]{@{}l@{}} S. Aslam, S. Munawar, M. Sumal, Z. Khan, M. W. Akram, The effect of humor orientation on team performance\\ with mediating role of
workplace bullying, The journal of contemporary issues in business and government\\ 27 (6) (2021) 159–170.(\cite{aslam2021effect})\end{tabular} \\

[22] & \begin{tabular}[c]{@{}l@{}} H. S. Al Obthani, R. B. Omar, et al., A contextual model on the role of management in\\ fostering humor at work, International Journal of
Business and Social Science 3 (24) (2012)(\cite{al2012contextual})\end{tabular} \\

[23] & \begin{tabular}[c]{@{}l@{}} K. N. Khan, M. F. Jan, K. Wazir, Z. Khan, M. Z. Safdar, Leading with happiness; analysing \\the effect of leaders’ humour styles on employees’
creativity., Webology 19 (3) (2022).(\cite{khan2022leading})\end{tabular} \\

[24] & \begin{tabular}[c]{@{}l@{}} W. G. Biemans, E. K. Huizingh, Why so serious? the effects of humour on creativity and \\innovation, Creativity and Innovation Management
33 (2) (2024) 181–196.(\cite{biemans2024so})\end{tabular} \\

[25] & \begin{tabular}[c]{@{}l@{}} T. H. Pham, L. K. Bartels, Laughing with you or laughing at you: The influence of playfulness and humor on \\employees’ meeting satisfaction
and effectiveness, Journal of Organizational Psychology\\ 21 (5) (2021) 1–18.(\cite{pham2021laughing})\end{tabular} \\

[26] & \begin{tabular}[c]{@{}l@{}}  M. G. R. Pandin, C. S. Waloejo, M. Munir, H. K. Dewi, The influence of humor on students resilience, \\resmilitaris 13 (3) (2023) 1683–1693(\cite{pandin2023influence})\end{tabular} \\

[27] & \begin{tabular}[c]{@{}l@{}} T. Liu, H. Yang, F.-J. Wang, A creative approach to humour degree calculation for utterances,\\ in: 2020 IEEE 20th International Conference
on Software Quality, Reliability and Security Companion\\ (QRS-C), IEEE, 2020, pp. 650–656.(\cite{liu2020creative})\end{tabular} \\

[28] & \begin{tabular}[c]{@{}l@{}} A. Amjed, S. H. S. Tirmzi, Effect of humor on employee creativity with moderating role of \\transformational leadership behavior, Journal of
Economics, Business and Management \\4 (10) (2016) 594–598.(\cite{amjed2016effect})\end{tabular} 
\\

\end{tabular}%
}
\end{table*}





\end{document}